\documentstyle[aps,prb,epsfig]{revtex}

\begin{document}

\draft

\title{Flux noise in high-temperature superconductors}
\author{Carsten Timm\cite{present}}
\address{Universit\"at Hamburg, I. Institut f\"ur Theoretische
Physik, Jungiusstrasse 9, D-20355 Hamburg, Germany}
\date{September 6, 1996}

\maketitle

\begin{abstract}
Spontaneously created vortex-antivortex pairs are the predominant
source of flux noise in high-temperature superconductors.
In principle, flux noise measurements
allow to check theoretical predictions for both the
distribution of vortex-pair sizes and for the vortex diffusivity.
In this paper the flux-noise power spectrum is calculated
for the highly anisotropic high-temperature superconductor
Bi$_2$Sr$_2$CaCu$_2$O$_{8+\delta}$, both for bulk crystals and
for ultra-thin films. The spectrum is basically given by
the Fourier transform of the temporal magnetic-field correlation
function. We start
from a Berezinskii-Kosterlitz-Thouless type theory and
incorporate vortex diffusion, intra-pair vortex interaction,
and annihilation of pairs by means of a Fokker-Planck equation
to determine the noise spectrum
below and above the superconducting transition temperature.
We find white noise at low
frequencies $\omega$ and a spectrum proportional to
$1/\omega^{3/2}$ at high frequencies. The cross-over frequency
between these regimes strongly depends on temperature.
The results are compared with
earlier results of computer simulations.
\end{abstract}


\widetext

\section{INTRODUCTION}
\label{sec:intro}

It is well known that the layered structure of cuprate
high-temperature superconductors (HTSC's) leads to enhanced
two-dimensional fluctuations.
These fluctuations are partly due to spontaneously created pancake
vortex pairs in the superconducting CuO$_2$ layers. There are
se\-ve\-ral
attempts\cite{layBKT} to describe these vortices starting from the
Berezinskii-Kosterlitz-Thouless (BKT) renormalization group
theory.\cite{BKT} These approaches differ in their
predictions so that experiments are needed to decide between them.

Most experiments designed to test the predictions
of BKT-type theories indirectly measure the temperature dependence of
the renormalized interaction. This quantity can be obtained from
the exponent $\alpha$ of the non-linear current-voltage
characteristics $V\propto I^\alpha$.\cite{HN,SPnew}
A second approach is to measure the linear resistivity above $T_c$,
which is related to the superconducting correlation length.
It has been shown, however,
that the derivation of the resistivity within the framework of BKT
theory is at best only valid in a narrow temperature range, which is
probably inaccessible experimentally.\cite{Mi:Nobel}
Thus, most of our experimental knowledge
about vortex pair fluctuations is based on measurements of the
temperature dependence of just one quantity.
Alternative approaches would be very welcome.

Apart from the renormalized interaction and the correlation length,
BKT-type theories also predict the tem\-pe\-ra\-ture- and
size-de\-pen\-dent fugacity of pairs, and, consequently, the
distribution of pair sizes and the
total pair density---at least below the transition temperature.
A generalized approach\cite{Ti:PhC}
yields quantitative results even above $T_c$.
It takes care of the correct counting of overlapping
vortex-antivortex pairs and takes local-field effects in the
screened interaction into account.
In this way terms of higher order in the vortex fugacity $y$
are introduced into the Kosterlitz recursion relations,\cite{BKT}
facilitating a description of the vortex system even above $T_c$.


Only few experiments sensitive to the pair density have been
performed, most of them on magnetic-flux noise.
In the absence of an external magnetic field, the flux at
the surface of an HTSC sample is due to the vortices in the bulk.
This flux is noisy since
these vortices perform a diffusive motion, carrying their magnetic
field with them. Only few of these experiments have been done on
HTSC's, mostly on bulk YBa$_2$Cu$_3$O$_{6+\delta}$
(Y-123).\cite{Ferr94} However, BKT-type theories and hence the
approach presented here are probably not applicable to Y-123 since
its anisotropy is too small. There are also noise measurements on
Josephson junction arrays.\cite{Shaw} These arrays are discrete
systems with a relatively large lattice constant, for which
the continuum approach presented below is not suitable.

Rogers {\it et al}.\cite{Rogers}\ perform
experiments on very thin films of Bi-2212 in the absence of
an external magnetic field. 
(Apparently experiments on bulk single crystals of Bi-2212
have not been performed yet.)
The authors find a flux-noise spectrum following
a $\omega^{-3/2}$ law for frequencies $\omega\gtrsim\omega_c$,
where the characteristic frequency
$\omega_c$ strongly increases with temperature.


In this paper we determine the effect of vortex-pair fluctuations
in both bulk HTSC's and ultra-thin films (containing one CuO$_2$
layer) on flux noise. To be specific,
we consider the highly anisotropic compound
Bi$_2$Sr$_2$CaCu$_2$O$_{8+\delta}$
(Bi-2212) in a vanishing external magnetic field under the assumption
that the superconducting layers are coupled only weakly
so that the dynamics of the vortices in one layer is independent of
that in the other layers. We further assume a large density of similar
pinning centers.
We will find that the spectral density of flux noise
is governed by the temporal magnetic-field correlation function.
Interestingly, the same correlation function also governs
the contribution of vortex-pair fluctuations to
nuclear-spin relaxation,\cite{ATZ,Diss} albeit at much higher
frequencies.

Ambegaokar {\it et al}.\cite{Ambe} employ a Fokker-Planck equation
to obtain the linear response of a superfluid film to substrate
oscillations. In this equation the authors include the
interaction between vortices within the same pair.
Similarly, we also start from a Fokker-Planck equation.
However, we solve this equation to obtain the full
space- and time-dependent vortex correlations needed for
the calculation of the magnetic-field correlation function.

A similar vortex system is studied by
Houlrik {\it et al}.\cite{Houl} They derive a
relation between the flux-noise power spectrum and the
dissipation due to the vortices described by a dielectric
constant $\epsilon$. This relation is valid in the limiting case of
a {\it large\/} pick-up coil, i.e., for the flux through a large area.
Houlrik {\it et al}.\cite{Houl} perform
computer si\-mu\-la\-tions on a generalized two-dimensional, discrete
{\it XY\/} model\cite{genXY} to obtain $\epsilon$. The
mentioned relation is employed to get the noise spectrum.
It falls off as $1/\omega^2$ at very high frequencies $\omega$
and shows a $1/\omega^{3/2}$ dependence for smaller $\omega$.
The $\omega^{-3/2}$ power law is in agreement with
Minnhagen's phenomenological approach.\cite{MP}
In the present paper results for a continuous two-dimensional
Coulomb gas model are obtained by direct calculation 
as opposed to simulations. Furthermore, we consider the
opposite limiting case of a {\it small\/} pick-up coil.


The present paper is organized as follows: In Sec.~\ref{sec:spec}
we define the flux noise power spectrum and express it in terms of a
magnetic-field correlation function. In Sec.~\ref{sec:model} we
present a model which enables us to calculate this function and in
Sec.~\ref{sec:res} we discuss our results.

\section{THE FLUX NOISE POWER SPECTRUM}
\label{sec:spec}

We have the following setup in mind: The small pick-up coil of
a SQUID (superconducting quantum interference device)
magnetometer is placed at the surface of a large HTSC single crystal
or of an extended one-unit-cell-thick film.
For epitactically grown samples, the most natural way to mount the
input coil is on a $\{001\}$ ($ab$) plane, sensitive to the field
perpendicular
to the layers. In the following we restrict ourselves to this case.
The flux signal is
measured in the absence of any external field or driving force. The
spectrum is then obtained by Fourier transformation.

The flux-noise power spectrum $S_\phi$ is given by
the Wiener-Khinchin theorem,\cite{vanK}
\begin{equation}
S_\phi(\omega) = \frac{2}{\pi} \int_0^\infty \!\! dt\;\overline{
  \phi(t)\phi(0)}\, \cos\omega t .
\end{equation}
Here, $\phi(t)$ is the flux through the effective area
$A_{\text{eff}}$ of the
input coil. If the diameter of the effective area is smaller than
the typical length scale of magnetic field changes, $\lambda_{ab}$,
the field is approximately uniform over the
area $A_{\text{eff}}$, and we can write
\begin{equation}
S_\phi(\omega) = \frac{2}{\pi} A_{\text{eff}}^2
  \int_0^\infty\!\! dt\,\tilde k_{zz}(t)
  \cos\omega t 
  \equiv \sqrt{\frac{2}{\pi}}\,A_{\text{eff}}^2\,
  \tilde k_{zz}(\omega) .
\label{Sbykzz}
\end{equation}
We have thus reduced the problem to the determination of the Fourier
transform of the magnetic-field correlation function
\begin{equation}
\tilde k_{zz}(t) = \overline{ h_{0,z}({\bf r},t)
  h_{0,z}({\bf r},0) } ,
\label{cor.kab}
\end{equation}
where $h_{n,z}({\bf r},t)$ is the $z$ component of the total magnetic
field at the point ${\bf r}$ in layer $n$ at the time $t$.

\section{MODEL}
\label{sec:model}

\subsection{General considerations}

In this section we present a model for the time-dependent
local magnetic field in a layered superconductor in the absence of an
external field. Results for a single layer are then obtained by means
of a straightforward generalization.
We assume that the Josephson coupling between the
superconducting layers can be neglected as far as the dynamics
of pancake vortices is concerned.
Then the local field is due to spontaneously created pancake vortices
in the layers. We assume that there are $N$ vortices and $N$
antivortices in each layer at any time, thus
neglecting fluctuations of the vortex number. This is
justified since we are only interested in the thermodynamic limit.

We decompose the vortex system into the smallest possible
vortex-antivortex pairs, using the enumeration algorithm given by
Halperin,\cite{Halp} i.e., we count the vortex and the antivortex
with the smallest separation as a pair and then repeat this step
for the remaining vortices and antivortices.
Let ${\bf H}_n({\bf r})$ be the magnetic field of a single vortex
situated at the origin in layer zero measured
at the point ${\bf r}$ in the $n$-th layer.
Here, we only need the $z$ component of the field.
It is given by\cite{Arte}
\begin{equation}
H_{n,z}({\bf r}) = \!\frac{\phi_0s}{4\pi\lambda_{ab}^2
  \sqrt{r^2+n^2s^2}}
  \exp\!\left(-\frac{\sqrt{r^2+n^2s^2}}{\lambda_{ab}}\right) ,
\label{def.Hr.3}
\end{equation}
where $\phi_0$ is the superconducting flux quantum and $s$ is the
interlayer spacing.
This expression holds for an infinite stack of superconducting
layers. The field differs from this result outside the crystal,
where it is not screened. However, if the pick-up coil is placed
close to the surface the difference should be negligible.
The two-di\-men\-sio\-nal symmetric Fourier transform of
Eq.~(\ref{def.Hr.3}) is
\begin{eqnarray}
H_{n,z}({\bf k}) & = & \frac{\phi_0s}{4\pi \lambda_{ab}^2}\,
  \frac{1}{\sqrt{ k^2+\lambda_{ab}^{-2}}}
  \exp\!\left(-|n|\,s \sqrt{k^2+\lambda_{ab}^{-2}}
  \right) . \nonumber \\
& & 
\label{def.Hk.3}
\end{eqnarray}
For now we only utilize the fact that the field of an
antivortex is just the negative of ${\bf H}_n({\bf r})$. The fields
of the vortices and antivortices are superposed to obtain the total
magnetic field ${\bf h}_n({\bf r},t)$, which depends on time only
through the positions of the vortices and antivortices.

We are interested in the correlation function $\tilde k_{zz}(t)$
as given by Eq.~(\ref{cor.kab}).
The total magnetic field in layer zero is
\begin{eqnarray}
{\bf h}_0({\bf r},t) & = & \sum_{n=1}^\infty \sum_{\nu=1}^N \Big(
   {\bf H}_n[{\bf r}-{\bf r}_{-n,\nu +}(t)] \nonumber \\
& & {}- {\bf H}_n[{\bf r}-{\bf r}_{-n,\nu -}(t)] \Big) ,
\end{eqnarray}
where ${\bf r}_{n,\nu +}(t)$ (${\bf r}_{n,\nu -}(t)$) is
the position of the
vortex (antivortex) of the $\nu$-th pair in layer $n$ at the time $t$.
We now assume that inter-pair correlations are negligible as compared
with intra-pair correlations. We keep the correlations between the
fields of the vortex and the antivortex of the same pair, however.
This approximation is justified if the typical
pair size is small as compared with the average distance between
neighboring pairs. Under the same condition the (extended) BKT
theory\cite{Ti:PhC} is applicable. If we further assume diffusive
dynamics we can write down the following ansatz for the correlation
function:\cite{ATZ}
\begin{eqnarray}
\tilde k_{zz}(t) & = & \frac{2N}{F} \sum_{n=1}^\infty
  \int d^2r_+'d^2r_-'d^2r_+d^2r_-\, \nonumber \\
& & \times
  \big[ H_{n,z}({\bf r}-{\bf r}_+')\, H_{n,z}({\bf r}-{\bf r}_+)
  \nonumber \\
& & \quad
  {}- H_{n,z}({\bf r}-{\bf r}_+')\, H_{n,z}({\bf r}-{\bf r}_-) \big]
  \nonumber \\
& & \times P({\bf r}_+',{\bf r}_-';{\bf r}_+,{\bf r}_-;t)\,
  f({\bf r}_+-{\bf r}_-) . 
\label{cor.k5}
\end{eqnarray}
Here, $f({\bf r})$ is the normalized distribution function of the
pair separation vector ${\bf r}={\bf r}_+-{\bf r}_-$. We obtain this
function numerically from the extended BKT theory of
Refs.~\onlinecite{Ti:PhC} and \onlinecite{Diss}. For our calculations
we use an approximate form of $f({\bf r})$, which incorporates the
essential physics, cf.\ Subsec.~\ref{sus:f}. The diffusive motion of
the pairs is described by the time-evolution kernel or diffusion
function $P$:
$P({\bf r}_+',{\bf r}_-';{\bf r}_+,{\bf r}_-;t)\, d^2r_+' d^2r_-'$
is the probability of finding
the vortex of a given pair in the area element $d^2r_+'$
about ${\bf r}_+'$ and the antivortex of the same pair
in $d^2r_-'$ about ${\bf r}_-'$ at the time $t$ provided that the
vortex was at ${\bf r}_+$ and the antivortex at ${\bf r}_-$ at the
time zero.
The indices $n$ and $\nu$ of the vortex positions have been omitted
in Eq.~(\ref{cor.k5}) since we are dealing with one representative
pair.

\subsection{The diffusion function}

It may be instructive to turn briefly to the case of unbound
pairs. In this case the vortices diffuse independently and
the diffusion function separates into
a product of free diffusion functions for the two partners of the
pair,
\begin{eqnarray}
P({\bf r}_+',{\bf r}_-';{\bf r}_+,{\bf r}_-;t)
  & = & \frac{1}{4\pi Dt}\,
       \exp\!\left(-\frac{|{\bf r}_+'-{\bf r}_+|^2}{4Dt}\right)\;
  \nonumber \\
& & {}\times\frac{1}{4\pi Dt}\,
       \exp\!\left(-\frac{|{\bf r}_-'-{\bf r}_-|^2}{4Dt}\right)\! ,
  \!\!\!\!
\label{cor.P5}
\end{eqnarray}
where $D$ is the diffusion constant of a free vortex.
By rewriting Eq.~(\ref{cor.P5}) in terms of center-of-mass
and relative coordinates, we can
see that the center of mass diffuses freely with the diffusion
constant $D_{\text{CM}} = D/2$,
whereas the separation vector diffuses with $D_{\text{rel}} = 2D$.

Now we wish to take two important effects into account, namely
the interaction between the two partners of a pair and
the recombination of pairs. The latter effect is expected to
destroy the correlation on the time scale of the
recombination time. The
center of mass of the pair should perform a free diffusion.
The time-evolution kernel can then be written as
\begin{eqnarray}
\lefteqn{
P\left({\bf R}'+\frac{{\bf r}'}{2},{\bf R}'-\frac{{\bf r}'}{2};
  {\bf R}+\frac{{\bf r}}{2},{\bf R}-\frac{{\bf r}}{2};t \right) }
  \nonumber \\
& & = \frac{1}{2\pi Dt}\,\exp\!\left(-\frac{|{\bf R}'-{\bf R}|^2}{2Dt}
  \right) \:P_{\text{rel}}({\bf r}',{\bf r};t) .
\label{cor.Ptot}
\end{eqnarray}
The task at hand is to determine the time evolution of the
separation vector, $P_{\text{rel}}({\bf r}',{\bf r};t)$.

To obtain $P_{\text{rel}}$ we solve a Fokker-Planck equation
containing the intra-pair interaction $V$.\cite{Ambe}
The vortex-antivortex interaction is given by\cite{BKT}
\begin{equation}
V(r) = \int_{r_0}^r \!\! dr'\,\frac{q^2}{\epsilon(r')r'} .
\end{equation}
For most pairs the dielectric constant $\epsilon(r)$ is close to
unity\cite{Ti:PhC} so that we may replace $q^2/\epsilon$ by $q^2$
and write $V(r) \approx q^2\,\ln(r/r_0)$. 
The error thereby incurred turns out to be small as compared with
errors due to, e.g., the uncertainty of the
diffusion constant. Our approximation
is best justified for small pairs, for which the interaction
is screened only weakly.
For temperatures significantly above $T_c$ many large pairs with
strongly screened interaction exist and the approximation breaks down,
while the extended BKT theory also becomes invalid.

If the mobility and diffusivity are isotropic and constant in space
and time, the diffusion (Fokker-Planck) equation in the presence of
a potential $V$ reads\cite{vanK}
\begin{equation}
\frac{\partial P_{\text{rel}}}{\partial t}
  = \mu_{\text{rel}} P_{\text{rel}}\Delta V
  + \mu_{\text{rel}} ({\mbox{\boldmath $\nabla$}} V)
  \cdot{\mbox{\boldmath $\nabla$}} P_{\text{rel}}
  + D_{\text{rel}} \Delta P_{\text{rel}} ,
\end{equation}
where the mobility $\mu_{\text{rel}}$ is related to the diffusion
constant through the Einstein relation
$\mu_{\text{rel}}=D_{\text{rel}}/k_BT$. The initial condition is
$P_{\text{rel}}({\bf r},{\bf r}_0;0) = \delta({\bf r}-{\bf r}_0)$.
Inserting the logarithmic potential we find
\begin{equation}
\frac{\partial P_{\text{rel}}}{\partial t}
  = 2\pi\mu_{\text{rel}} q^2 \delta({\bf r}) P_{\text{rel}}
  + \mu_{\text{rel}} q^2 \frac{{\bf r}}{r^2}
    \cdot {\mbox{\boldmath $\nabla$}} P_{\text{rel}}
  + D_{\text{rel}}\Delta P_{\text{rel}} .
\label{diffeq2}
\end{equation}
The first term on the right-hand side contains a delta function.
This term yields a positive contribution to the time
derivative only at ${\bf r}=0$. Therefore, it causes a
$\delta$-function term to appear in $P$ at ${\bf r}=0$.
Such a contribution does not affect $P$ for ${\bf r}\neq 0$.
Since ``pairs'' with ${\bf r}=0$ are recombined and
do not contribute to the magnetic
field, we may omit the first term in Eq.~(\ref{diffeq2}).
After introduction of polar coordinates, the diffusion equation
can be solved by means of a separation ansatz,\cite{Diss}
\begin{eqnarray}
P_{\text{rel}}({\bf r}',{\bf r};t)
  & = & \frac{1}{4\pi D_{\text{rel}} t}
  \left(\frac{r'}{r}\right)^{\!\gamma}
  \exp\left(-\frac{r'^2+r^2}{4D_{\text{rel}} t}\right)
  \nonumber \\
& \times & \sum_{m=-\infty}^\infty e^{im(\varphi'-\varphi)}
  I_{\sqrt{\gamma^2+m^2}}
    \left(\frac{r r'}{2D_{\text{rel}} t}\right)\! ,
\label{Prel.expl}
\end{eqnarray}
where
\begin{equation}
\gamma = \frac{k_BT-q^2}{2k_BT} .
\label{gamma.th}
\end{equation}
and $I_\alpha$ is a modified Bessel function. The full
time-evolution kernel is obtained by inserting the solution for
$P_{\text{rel}}$ into Eq.~(\ref{cor.Ptot}).

For $t=0$ the diffusion function $P_{\text{rel}}$
is normalized to unity by
construction. At later times more and more weight is expected
to accumulate in the $\delta$-term
at ${\bf r}=0$ while the overall norm remains constant.
The weight outside of the central singularity
is obtained by integration over two-dimensional space,
\begin{equation}
||P_{\text{rel}}|| = \frac{\displaystyle
  \tilde\gamma\left(-\gamma,\frac{r^2}{4D_{\text{rel}}\,t}\right)}
  {\Gamma(-\gamma)} ,
\label{PrelN}
\end{equation}
where
\begin{equation}
\tilde\gamma(a,x) \equiv \int_0^x\!\! dt\,e^{-t} t^{a-1}
\end{equation}
is the incomplete gamma function.\cite{AS}

\begin{figure}
\centerline{
\epsfig{file=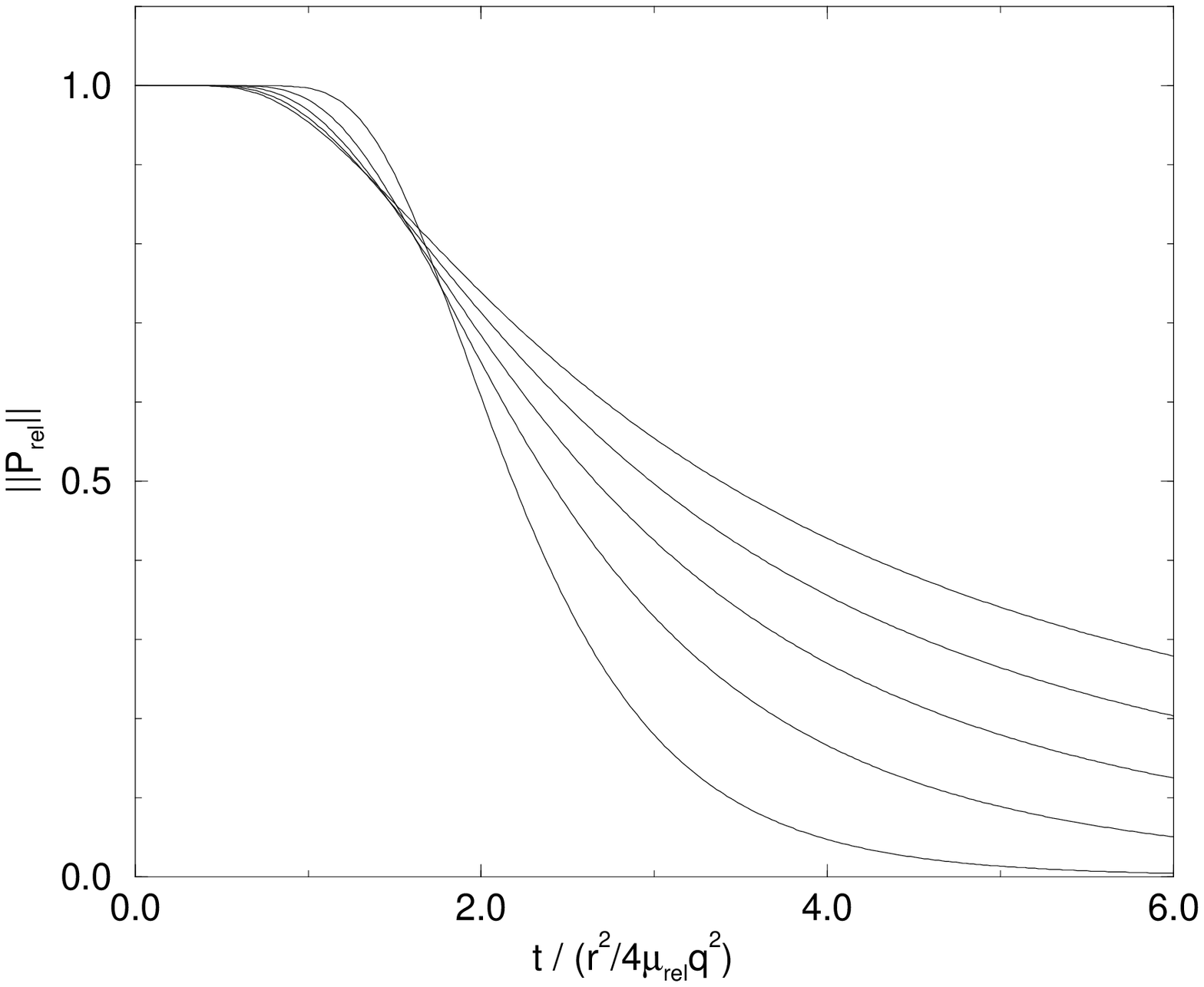,width=4.5in}}
\caption{The weight $||P_{\protect\text{rel}}||$
of the diffusion function outside the central singularity as a
function of time for temperatures
$k_BT/q^2=0.05, 0.1, 0.15, 0.2, 0.25$
(the steepest curve corresponds to $k_BT/q^2=0.05$). Time is measured
in units of $r^2/4\mu_{\protect\text{rel}} q^2$.}
\label{fig:PrelN}
\end{figure}

The expression (\ref{PrelN}) indeed approaches unity for $t\to0$, but
decreases monotonically with time and goes to zero for $t\to\infty$.
In particular, it behaves as $||P_{\text{rel}}|| \propto t^\gamma$
for large $t$ (note that $\gamma\le-3/2$). In
Fig.~\ref{fig:PrelN} the weight $||P_{\text{rel}}||$ is depicted
as a function of
time for various temperatures. The time is given in units of
$r^2/4\mu_{\text{rel}} q^2$
so that the curves are invariant under change
of $\mu_{\text{rel}}$.
The mobility $\mu_{\text{rel}}$ is kept constant.

As shown in Fig.~\ref{fig:PrelN},
there is a plateau in $||P_{\text{rel}}||$ for
small times and a sharp drop in the vicinity of an {\it annihilation
time\/} $\tau_a=r^2/4\mu_{\text{rel}} q^2$.
This is the typical time the separation vector
needs to diffuse from its initial value ${\bf r}$ to zero.
The curve $||P_{\text{rel}}||(t)$ is smeared out
at higher temperatures.
If the separation vector assumes the value ${\bf r}'=0$,
the pair is trapped by the singularity. Then the magnetic fields
cancel exactly and the pair has annihilated.
For low temperatures the pairs tend to creep ``downhill'' into the
potential well until they
annihilate after a time of the order of $\tau_a$.
At higher temperatures the diffusive motion
is generally faster so that the {\it first\/} pairs recombine earlier,
but many pairs first start to grow and recombine later.

Note that pairs are created at the same rate as they are destroyed.
However, newly created pairs do not contribute to the correlation
function since their positions are not correlated with the pairs
still existing or already destroyed.


\subsection{Distribution of pair sizes}
\label{sus:f}

Apart from the diffusion function, we also need to know the
distribution function of the separation vector, $f({\bf r})$, to
calculate the correlation function. Unfortunately the pair size
distribution is known only numerically.

In the direction parallel to the layers the magnetic field of a vortex
changes on a length scale given by the pe\-ne\-tra\-tion depth
$\lambda_{ab}$. Thus, the fields of a vortex and an antivortex with a
separation much smaller than $\lambda_{ab}$ almost cancel each other.
These small pairs do not contribute significantly to the correlation
function. We utilize this observation by approximating the pair size
distribution by an analytical expression which becomes exact for
large pairs.
The pair size distribution is intimately related to the pair fugacity
$y^2$ of BKT theory, $f({\bf r})=y^2(r)/r^4$. The modified
Kosterlitz recursion relations of the extended
BKT theory\cite{Ti:PhC,Diss} predict that $y^2$ and the renormalized
interaction described by the stiffness constant $K$ approach a finite,
temperature-dependend
fixed point $y^2(\infty)$, $K(\infty)$ for large length scales.
Hence, we can solve the recursion relations close to the
appropriate fixed point to obtain the leading behavior of the
fugacity, and thus of the pair size distribution $f({\bf r})$, at
large length scales. We find that $f({\bf r}) \propto 1/r^{2\zeta+4}$
with
\begin{equation}
\zeta = -2+\pi K(\infty)+2\pi^2 y^2(\infty) .
\end{equation}
From Ref.~\onlinecite{Ti:PhC} we see that the exponent $\zeta$
vanishes for $T\ge T_c$ and is positive and, to leading order,
proportional to $\sqrt{T_c-T}$ below $T_c$. Details may be found
in Ref.~\onlinecite{Diss}.

A reasonable approximation for the pair size distribution function
is
\begin{equation}
f({\bf r}) \propto \frac{1-(r/r_0)^2}
  {1-(r/r_0)^{2\zeta+6}} .
\end{equation}
This function
shows the correct behavior for large $r$ and does not introduce
irrelevant problems at small $r$.
Since BKT theory neglects pairs of size $r<r_0$, they are not
counted in the total density $N/F$.
The correct normalized distribution then reads
\begin{eqnarray}
f({\bf r}) & = & \frac{2\zeta+6}{2\pi r_0^2}\,
  \frac{1}{\Psi\!\left[1-2/(2\zeta+6)\right]
  - \Psi\!\left[1-4/(2\zeta+6)\right]} \nonumber \\
& & {}\times \frac{1-(r/r_0)^2}{1-(r/r_0)^{2\zeta+6}} ,
\end{eqnarray}
where $\Psi(x) = \Gamma'(x)/\Gamma(x)$ is the digamma
function.\cite{AS}

\subsection{Correlation functions}
\label{sus:corf}

Now we have all ingredients to calculate the
correlation function $\tilde k_{zz}$. Equation (\ref{cor.k5})
can be rewritten as
\begin{eqnarray}
\tilde k_{zz}(t) & = & \frac{2N}{F} \sum_{n=1}^\infty
  \int d^2R'd^2r'd^2R\,d^2r
  \bigg[ H_{n,z}\left({\bf R}'+\frac{{\bf r}'}{2}\right)
      H_{n,z}\left({\bf R}+\frac{{\bf r}}{2}\right) \nonumber \\
& & {}- H_{n,z}\left({\bf R}'+\frac{{\bf r}'}{2}\right)
      H_{n,z}\left({\bf R}-\frac{{\bf r}}{2}\right) \bigg]
  P_{\text{CM}}^{(0)}({\bf R}'-{\bf R};t)
  P_{\text{rel}}({\bf r}',{\bf r};t)
  f({\bf r}) ,
\end{eqnarray}
where $P_{\text{CM}}^{(0)}$ is the free diffusion function
of the center of mass.
To make this expression tractable numerically, we have to
analytically evaluate as many integrals as possible.
As noted above we need the temporal Fourier transform of the
correlation function. With $P_{\text{rel}}$ from
Eq.~(\ref{Prel.expl}) we get, as shown in Ref.~\onlinecite{Diss},
\begin{eqnarray}
\tilde k_{zz}(\omega) & = & \frac{2N}{F}\,
  \frac{8\sqrt{2\pi}}{D_{\text{rel}}}
  \int d^2k\,\sum_{n=1}^\infty\,|H_{n,z}({\bf k})|^2
  \int_0^\infty \!\! dr\,r^{1-\gamma} f(r)
  \sum_{m=1,\:m\text{ odd}}^\infty \!
  J_m\! \left(\frac{kr}{2}\right) \nonumber \\
& & \times \int_0^\infty \!\!
  dr'\,r'^{1+\gamma} J_m\!\left(\frac{kr'}{2}\right) \,
  \mbox{Re}\, I_{\sqrt{\gamma^2+m^2}}\!
  \left(\sqrt{\frac{k^2}{4} + i\frac{\omega}{D_{\text{rel}}}}
  \:r_<\right)
  K_{\sqrt{\gamma^2+m^2}}\!
  \left(\sqrt{\frac{k^2}{4} + i\frac{\omega}{D_{\text{rel}}}}
  \:r_>\right) ,
\end{eqnarray}
where $r_< = \min(r,r')$ and $r_> = \max(r,r')$. Taking into account
the special form of the vortex field as given by Eq.~(\ref{def.Hk.3}),
summing $|H_{n,z}|^2$ over the layers, and
performing the integral over the polar angle of ${\bf k}$ we get
\begin{eqnarray}
\tilde k_{zz}(\omega) & = & \frac{2N}{F}\,
  \frac{8\sqrt{2\pi}}{D_{\text{rel}}}\,
  \frac{\phi_0^2s^2}{8\pi\lambda_{ab}^4}
  \int_0^\infty \frac{dk\,k}{k^2+\lambda_{ab}^{-2}}\,
  \frac{1}{\exp\!\left(2s\sqrt{k^2+\lambda_{ab}^{-2}}\right)-1}
  \int_0^\infty \!\! dr\,r^{1-\gamma} f(r)
  \sum_{m=1,\:m\text{ odd}}^\infty \!
  J_m\! \left(\frac{kr}{2}\right) \nonumber \\
& & \times \int_0^\infty \!\!
  dr'\,r'^{1+\gamma} J_m\!\left(\frac{kr'}{2}\right) \,
  \mbox{Re}\, I_{\sqrt{\gamma^2+m^2}}\!
  \left(\sqrt{\frac{k^2}{4} + i\frac{\omega}{D_{\text{rel}}}}
  \:r_<\right)
  K_{\sqrt{\gamma^2+m^2}}\!\left(\sqrt{\frac{k^2}{4} +
  i\frac{\omega}{D_{\text{rel}}}}\:r_>\right) .
\label{kzz}
\end{eqnarray}
To describe a single layer we just have to replace the sum over $n$
by one term, say for $n=1$. This simply leads to the replacement
of $1/(\exp[2s(k^2+\lambda_{ab}^{-2})^{1/2}]-1)$
by $1/\exp[2s(k^2+\lambda_{ab}^{-2})^{1/2}]$ in Eq.~(\ref{kzz}).

Equation (\ref{kzz}) suggests that
$\omega_c\sim D_{\text{rel}}/4\lambda_{ab}^2$ is a characteristic
frequency of the correlation function since $\omega$ only
appears in the expression $k^2/4+i\omega/D_{\text{rel}}$ and
the characteristic value of $k$ is $1/\lambda_{ab}$ because
of the exponential. In fact $\lambda_{ab}$ is the largest
length scale in the problem so that $D_{\text{rel}}/4\lambda_{ab}^2$
is the smallest frequency where we expect the spectrum to show
any feature.

Of the parameters appearing in the rates the numerical value of
the diffusion constant $D_{\text{rel}}=2D$ is least well known.
Here, we briefly discuss vortex diffusion and its relation to pinning.
In the absence of pinning the friction coefficient $\eta$
of a vortex can be obtained from Bardeen-Stephen theory,\cite{BS}
$\eta = \phi_0^2/2\pi c^2\xi^2\rho_n$.
To take the anisotropy into account, one replaces $\xi$ by
the coherence length within the layers, $\xi_{ab}$.
We thus have\cite{Blatt}
$\eta_{ab} = \phi_0^2/2\pi c^2\xi_{ab}^2\rho_n = \epsilon\eta$
with the effective mass ratio $\epsilon^2=m/M<1$.
The mobility $\mu$ of a vortex is then
$\mu = 1/\eta_{ab}d$, where $d$ is the thickness of the
superconducting layers.
The diffusion constant is obtained using the Einstein relation,
\begin{equation}
D_0 = \mu k_BT = \frac{2\pi c^2\xi_{ab}^2(T)\rho_n k_BT}{\phi_0^2 d} .
\end{equation}
If one employs the Bardeen-Stephen formula the diffusion constant in
Bi-2212 turns~out to~be of the~order of $1\mbox{ cm}^2/\mbox{s}$,
which seems rather large.
However, since Bardeen-Stephen theory neither takes into account
the discreteness of the quasi-particle spectrum in the vortex cores
nor the apparent d-wave symmetry of the energy gap, it may well give
incorrect results for HTSC's.
Measurements of $D$ for HTSC's do not present a
consistent picture.\cite{FOG}
Diffusivities from $10^{-4}\mbox{ cm}^2/\mbox{s}$ to
$10^2\mbox{ cm}^2/\mbox{s}$ have been reported.

A large density of similar pinning centers leads to a thermally
activated behavior of the diffusion constant,\cite{Fish80}
\begin{equation}
D = D_0\exp\!\left(-\frac{E_p}{k_BT}\right) ,
\label{DT.1}
\end{equation}
where $E_p$ is the typical depinning energy.
Matters are complicated by the observation that the depinning energy
depends on temperature. Rogers {\it et al}.\cite{Rogers}\ find the
following empirical relation for Bi-2212 films:
\begin{equation}
E_p(T) \approx E_p^0\left(1-\frac{T}{T_{c0}}\right)
\label{DT.2}
\end{equation}
with $E_p^0/k_B \approx 1200\mbox{ K}$. Other experiments also
support a large value of the activation energy.\cite{Grish}
These results only hold on time scales longer than the typical
depinning time. For shorter times the description by means of
a diffusion equation breaks down and has to be replaced by a model
explicitly incorporating discrete hopping.

\section{RESULTS AND DISCUSSION}
\label{sec:res}

From Eq.~(\ref{Sbykzz}) and the correlation function given
by Eq.~(\ref{kzz})
we immediately obtain the noise spectrum
$S_\phi = \sqrt{2/\pi}\,A_{\text{eff}}^2\,\tilde k_{zz}(\omega)$.

For the numerical evaluation of $S_\phi$ we employ Monte-Carlo
integration. For each set of parameters and each value of the sum
index $m=1,3,\ldots$ we have performed $3$ to $40$ Monte-Carlo runs
with $5000$ sample points
each. We use the distribution of the results of the individual
runs to estimate the numerical error.
We find that the summands fall off quickly for increasing $m$
so that the term for $m=5$ is smaller than the error of the $m=1$
term. Hence, terms for $m=7,9,\ldots$ are neglected.

We first consider bulk Bi-2212.
For the Ginzburg-Landau coherence length and the magnetic
penetration depth we set $\xi_{ab}(T=0)\approx 21.5\mbox{ \AA}$ and
$\lambda_{ab}(T=0)\approx 2000\mbox{ \AA}$. To obtain the
lengths at a temperature $T$ we employ the Ginzburg-Landau formula
$\lambda_{ab}(T)/\lambda_{ab}(0) = \xi_{ab}(T)/\xi_{ab}(0)
= \sqrt{T_{c0}/(T_{c0}-T)}$, where $T_{c0}$ is the
mean-field transition temperature.
The density of vortices and the parameter $\zeta$
are determined from the extended two-dimensional
BKT theory of Ref.~\onlinecite{Ti:PhC}.
This is consistent since we have neglected
interlayer vortex correlations throughout this paper.
As noted above, the extended BKT theory\cite{Ti:PhC,Diss}
should be applicable even in a temperature interval above $T_c$.
For higher
temperatures, however, any description in terms of vortex pairs fails
and a picture of free vortices is more appropriate. In this case we
expect the spectrum to fall off as $1/\omega^2$.
The pa\-ra\-me\-ter $\gamma$ is given by Eq.~(\ref{gamma.th}). For the
coupling constant $q^2$ we make the
standard linear approximation\cite{Mooij} $q^2=q_0^2k_B(T_{c0}-T)$,
where $q_0^2$ can be obtained from the known values
of $k_BT_c/q^2(T_c)\approx 0.2053$,
$T_c\approx 84.7\mbox{ K}$, and
$T_{c0}\approx 86.8\mbox{ K}$.\cite{Mart}

\begin{figure}
\centerline{
\epsfig{file=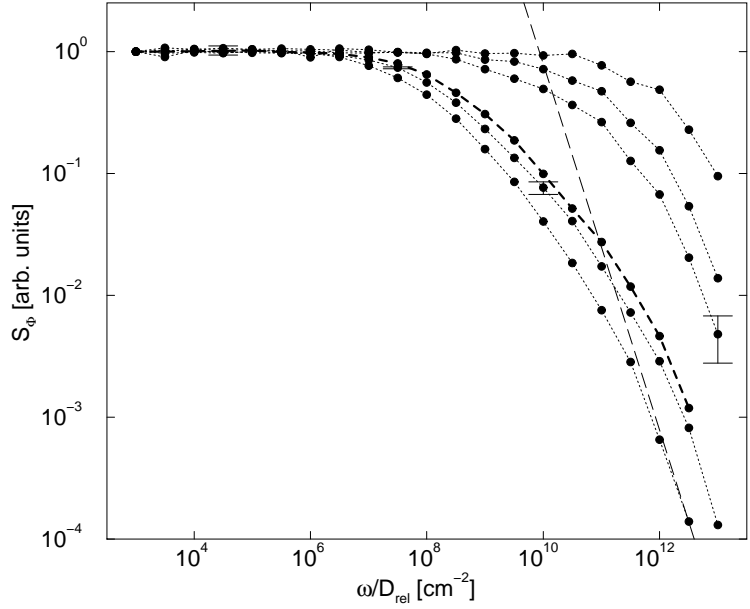,width=4.5in}}
\caption{Double-logarithmic plot of the flux-noise power spectrum
for a $c$ axis oriented Bi-2212 bulk single
crystal at $T=84.2, 84.5, 84.6, 84.7, 85.2, 85.7\mbox{ K}$
(from top to bottom). The critical temperature is $T_c=84.7\mbox{ K}$.
The dashed line corresponds to the power law
$S_\phi\propto \omega^{-3/2}$. Exemplary error bars are also shown.}
\label{fig:noise}
\end{figure}

Since the diffusivity is not well known, we first display
$S_\phi$ in arbitrary units at six different temperatures
in the vicinity of $T_c$
as a function of $\omega/D_{\text{rel}}$ in Fig.~\ref{fig:noise}.
Exemplary error bars from Monte-Carlo integration are also shown.
Displayed in this way the curves do not depend on $D_{\text{rel}}$.
The units of $S_\phi$ are chosen in such a way that
$S_\phi=1$ for $\omega/D_{\text{rel}}=10^3\mbox{ cm}^{-2}$.
The absolute value of the
noise power is thus not comparable for different temperatures.
Because of this choice of units, the pair density,
which is a simple factor in $S_\phi$, does not enter the
calculation. The calculation thus becomes independent of
$D_{\text{rel}}$ and $N/F$, which are the two most uncertain
quantities.

The spectra show a cross-over from 
white noise at low frequencies to a drop at higher frequencies.
The drop is found to approach the power law
$S_\phi\propto\omega^{-3/2}$
(the dashed line in the figure). The same behavior is found by
Houlrik {\it et al}.\cite{Houl} in their simulations, except at
very high frequencies. A $\omega^{-2}$ drop in that regime,
as seen by Houlrik {\it et al.}, is not found. However, we cannot
investigate higher frequencies since the numerical errors start
to increase rapidly. Since higher frequencies correspond to
shorter probed length scales the vortices should eventually
act as free particles, leading to a $\omega^{-2}$ power law.

The two frequency regimes of white noise and a $\omega^{-3/2}$
power law are separated by a characteristic value
of $\omega/D_{\text{rel}}$. Just below $T_c$,
this characteristic value strongly decreases with increasing
temperature, whereas the temperature dependence is weaker for
$T\ge T_c$. Since the only quantity in our calculations that shows
a similar
behavior is the exponent $\zeta$ of the distribution function $f$,
the main source of the temperature dependence of $\omega_c$
has to be $\zeta$.  A more rapid drop of $f({\bf r})$, corresponding
to smaller average pair size, leads to shorter recombination times
and thus to higher characteristic frequencies.
In this way measurements of $S_\phi(\omega)$ probe
the distribution of pair sizes.
For $T\gtrsim T_c$ the characteristic frequency
is of the order of $10^7\mbox{ cm}^{-2}\times D_{\text{rel}}$,
corresponding to a characteristic length
$3\times 10^{-4}\mbox{ cm}$, which is indeed of the order of
$\lambda_{ab}(T_c)$. (Note that $\lambda_{ab}$ diverges
at $T_{c0}>T_c$.)

\begin{figure}
\centerline{
\epsfig{file=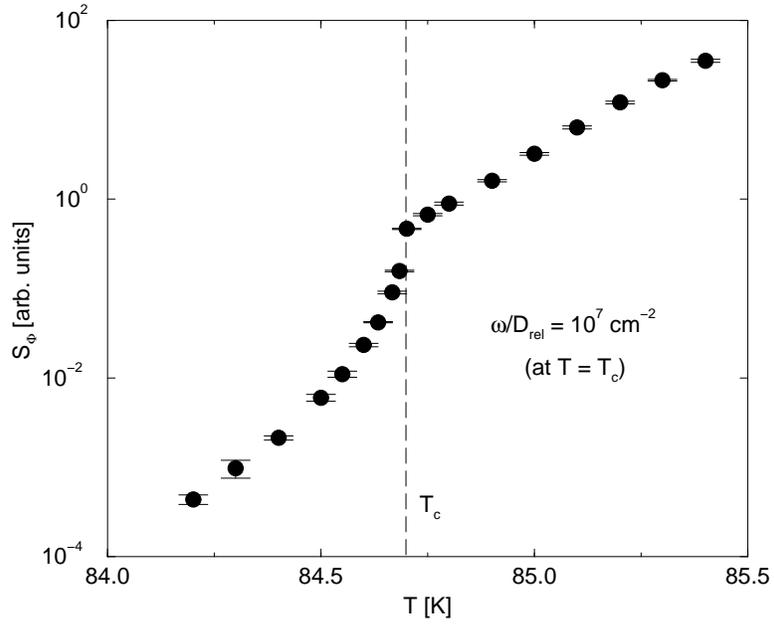,width=4.5in}}
\caption{The flux noise power for a $c$ axis oriented Bi-2212 crystal
as a function of temperature at fixed frequency.
The error bars from Monte-Carlo integration are also shown.}
\label{fig:soT}
\end{figure}

To be able to compare the flux noise at different temperatures, we
have to take the temperature dependence of both the prefactor of
$S_\phi$ and the diffusion constant $D_{\text{rel}}$ into
account. As a result we show the absolute noise power for bulk Bi-2212
at constant frequency as a function of temperature
in Fig.~\ref{fig:soT}. For any temperature,
the value of $\omega/D_{\text{rel}}$ is fixed by the requirement
that $\omega/D_{\text{rel}}=10^7\mbox{ cm}^{-2}$ at $T=T_c$,
together with the known temperature dependence of $D_{\text{rel}}$,
cf.\ Eqs.~(\ref{DT.1}) and (\ref{DT.2}).
The noise strongly increases with temperature, which we
mainly attribute to the temperature-dependence of the vortex
density $2N/F$. The density increases approximately exponentially
as more and more pairs are thermally excited.\cite{Ti:PhC}
Additionally, there is a kink at $T_c$, which should be the result of
the kink in the exponent $\zeta$. Since flux noise is
dominated by large pairs, the increasing exponent $\zeta$ below
$T_c$ leads to an additional reduction of the noise.
The characteristic form of $S_\phi(T)$ shown in Fig.~\ref{fig:soT}
can serve as an indication of a BKT-type transition.

With the diffusion constant $D$ determined from Bar\-deen-Ste\-phen
theory,\cite{BS} the characteristic frequency lies outside the
experimentally accessible frequency range.\cite{Ferr94}
We have argued above, however, that Bardeen-Stephen theory may be
unapplicable to HTSC's.
Turning the argument around, one could determine $D_{\text{rel}}$
from experimental spectra. Hence, experiments on bulk samples close to
$T_c$ are called for.
Voss and Clarke\cite{VC} argue that a spectrum
with $S_\phi\propto 1/\omega^{3/2}$ is expected for
$\omega \gtrsim 2D/A_{\text{eff}}$ due to diffusion of
vortices out of the sampled area. Since we
consider the case of a very small pick-up coil, the
cross-over to $1/\omega^{3/2}$ is expected to take place at
rather high frequencies. Thus the high cross-over frequencies
may be the result of our assumption of a small coil.

\begin{figure}
\centerline{
\epsfig{file=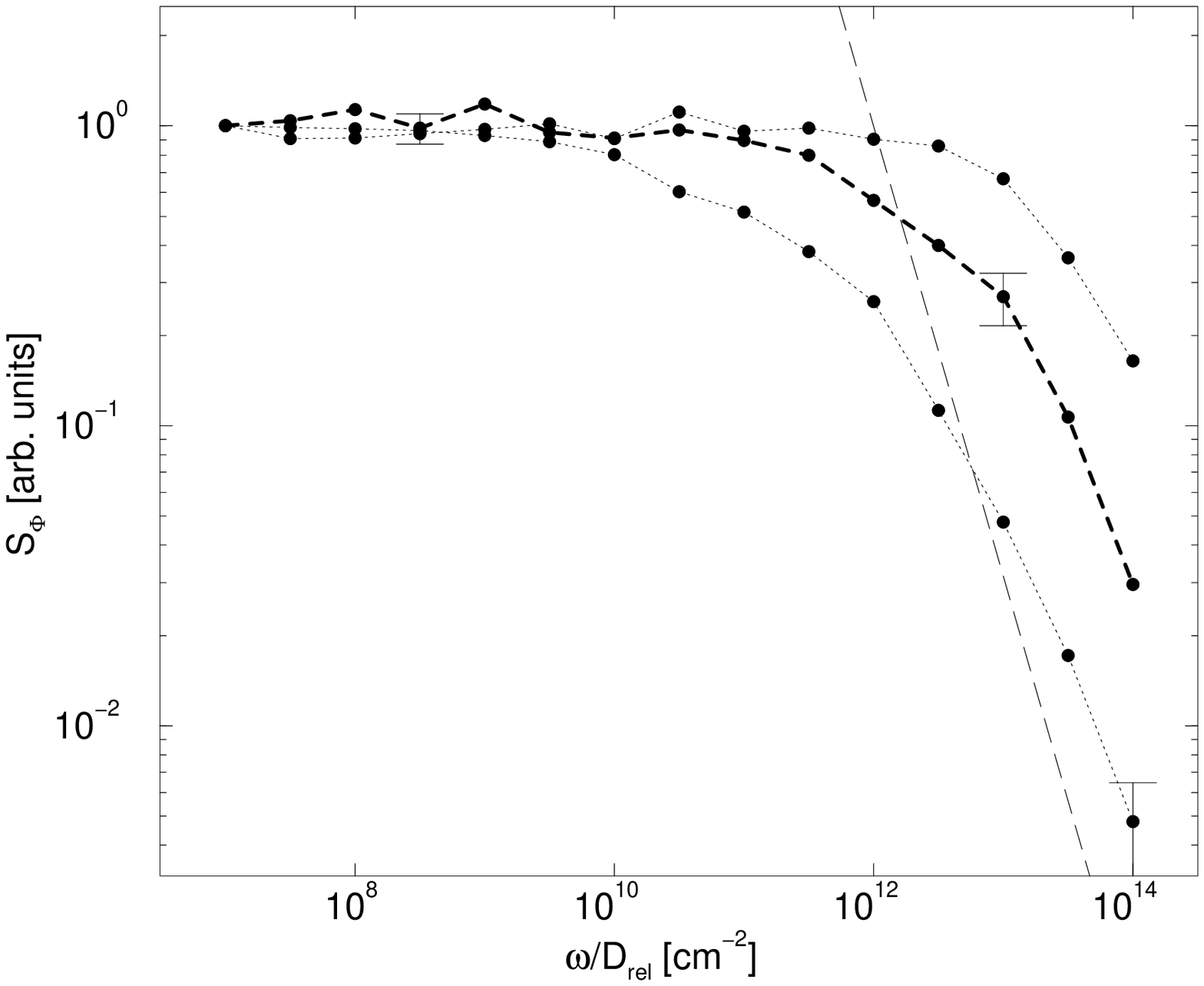,width=4.5in}}
\caption{Log-log plot of the flux-noise power spectrum for
a ultra-thin Bi-2212 film at $T=28, 31, 34\mbox{ K}$
(from top to bottom). The critical temperature if $T_c=31\mbox{ K}$.
The dashed line corresponds to $S_\phi\propto\omega^{-3/2}$.
Exemplary error bars are shown.}
\label{fig:film}
\end{figure}

We now turn to ultra-thin films of Bi-2212.
We use the same parameters as for bulk Bi-2212 with the exception
of the BKT temperature, which we choose as
$T_c\approx 31\mbox{ K}$ to allow comparison with the experiments
by Rogers {\it et al}.\cite{Rogers}
Figure \ref{fig:film} shows the flux noise spectrum of a film at
three different temperatures. The units are chosen as before.
Again, the spectra show white noise at low frequencies and a drop
at higher frequencies. The spectrum does not follow a $\omega^{-3/2}$
power law in the frequency range considered here. Weak convergence
of Monte Carlo data precludes calculations for higher frequencies.
However, we have no reason to doubt that $\omega^{-3/2}$ behavior is
eventually reached. The qualitative shape of the spectra
agrees with Ref.~\onlinecite{Rogers}.

The cross-over frequency is again found to decrease with temperature,
consistent with our picture of larger and larger pairs with
increasing temperature, which take longer to recombine.
This result is in contradiction to the experiments of Rogers
{\it et al}.\cite{Rogers} and the simulations of
Houlrik {\it et al}.\cite{Houl}, who find an increasing cross-over
frequency. These are results for the opposite limiting case
of a large pick-up loop, however.
The origin of the discrepancy is not yet clear.
Note that the simulations suggest a {\it vanishing\/} cross-over
frequency at $T=T_c$, which does not seem to be
consistent with experiment.

To conclude, we have used a model which is based on the assumption of
diffusing vortex-antivortex pairs and incorporates
intra-pair interaction and pair annihilation
to obtain flux-noise spectra for Bi-2212 single crystals and films.
The spectra
show white noise up to a strongly temperature-dependent cross-over
frequency and $1/\omega^{3/2}$ noise above. As a function of
temperature, the noise shows a distinct kink at the BKT temperature
$T_c$. We have shown that
flux noise measurements can yield information about the size
distribution of vortex-antivortex pairs and on vortex dynamics,
and can be used as an additional probe for a BKT
transition.

\acknowledgements

The author is obliged to J. Appel and T. \hbox{Wolenski} for
interesting discussions. Financial support by the Deutsche
Forschungsgemeinschaft (DFG) is gratefully acknowledged.


\newpage


\begin{references}
\bibitem[*]{present} Present address: Indiana University,
Dept.\ of Physics, Bloo\-ming\-ton, IN 47405.
\bibitem{layBKT}S. Scheidl and G. Hackenbroich, Europhys.\ Lett.\
{\bf 20}, 511 (1992);
B. Horovitz, Phys.\ Rev.\ B {\bf 47}, 5947 (1993);
G. Blatter, B.I. Ivlev, and H. Nordborg, Phys.\ Rev.\ B
{\bf 48}, 10~448 (1993);
K.H. Fischer, Physica C {\bf 210}, 179 (1993);
S.W. Pierson, Phys.\ Rev.\ Lett.\ {\bf 73}, 2496 (1994);
M. Friesen, Phys.\ Rev.\ B {\bf 51}, 632 (1995);
S.W. Pierson, Phys.\ Rev.\ B {\bf 51}, 6663 (1995);
C. Timm, Phys.\ Rev.\ B {\bf 52}, 9751 (1995).
\bibitem{BKT}V.L. Berezinskii, Zh.\ Eksp.\ Teor.\ Fiz.\ {\bf 61}, 1144
(1971) [Sov.\ Phys.\ JETP {\bf 34}, 610 (1972)];
J.M. Kosterlitz and D.J. Thouless, J.~Phys.\ C {\bf 6},
1181 (1973); J.M. Kosterlitz, J. Phys.\ C {\bf 7}, 1046 (1974).
\bibitem{HN}B.I. Halperin and D.R. Nelson, J. Low Temp.\ Phys.\
{\bf 36}, 599 (1979).
\bibitem{SPnew}S.W. Pierson, Phys.\ Rev.\ Lett.\ {\bf 74},
2359 (1995).
\bibitem{Mi:Nobel}P. Minnhagen and P. Olsson, Physica Scripta
{\bf T42}, 29 (1992).
\bibitem{Ti:PhC}C. Timm, Physica C {\bf 265}, 31 (1996).
\bibitem{Ferr94}M.J. Ferrari, M. Johnson, F.C. Wellstood, J.J.
Kingston, T.J. Shaw, and J. Clarke, J. Low Temp.\ Phys.\ {\bf 94}, 15
(1994).
\bibitem{Shaw}T.J. Shaw, M.J. Ferrari, L.L. Sohn, D.-H. Lee,
M. Tinkham, and J. Clarke, Phys.\ Rev.\ Lett.\ {\bf 76}, 2551 (1996).
\bibitem{Rogers}C.T. Rogers, K.E. Myers, J.N. Eckstein, and
I. Bozovic, Phys.\ Rev.\ Lett.\ {\bf 69}, 160 (1992).
\bibitem{ATZ}J. Appel, C. Timm, and A. Zabel, J. Low Temp.\ Phys.\
{\bf 99}, 553 (1995).
\bibitem{Diss}C. Timm, {\it Effects of Vortex Fluctuations in
High-Tem\-pe\-ra\-ture Superconductors}, thesis, Universit\"at Hamburg
(Sha\-ker, Aa\-chen, 1996, ISBN 3-8265-1652-4).
\bibitem{Ambe}V. Ambegaokar and S. Teitel, Phys.\ Rev.\ B {\bf 19},
1667 (1979); V. Ambegaokar, B.I. Halperin, D.R. Nelson, and
E.D. Siggia, Phys.\ Rev.\ B {\bf 21}, 1806 (1980).
\bibitem{Houl}J. Houlrik, A. Jonsson, and P. Minnhagen,
Phys.\ Rev.\ B {\bf 50}, 3953 (1994).
\bibitem{genXY}E. Domani, M. Schick, and R. Swendsen, Phys.\ Rev.\
Lett.\ {\bf 52}, 1535 (1984).
\bibitem{MP}P. Minnhagen, Rev.\ Mod.\ Phys.\ {\bf 59}, 1001 (1987).
\bibitem{vanK}N.G. van Kampen, {\it Stochastic Processes in
Physics and Chemistry\/} (North-Holland, Amsterdam, 1981).
\bibitem{Halp}B.I. Halperin, in {\it Proceedings of Kyoto Summer
Institute 1979---Physics of Low-Dimensional Systems}, edited by
Y. Nagaoka and S. Hikami (Publication Office, Prog.\ Theor.\ Phys.,
Kyoto), p.~53.
\bibitem{Arte}S.N. Artemenko and A.N. Kruglov, Phys.\ Lett.\ A
{\bf 143}, 485 (1990); M.V. Feigel'man, V.B. Geshkenbein, and A.I.
Larkin, Physica C {\bf 167}, 177 (1990); J.R. Clem, Phys.\ Rev.\ B
{\bf 43}, 7837 (1991).
\bibitem{AS}M. Abramovitz and I. Stegun, {Handbook of Mathematical
Functions\/} (Dover, New York, 1972).
\bibitem{BS}J. Bardeen and M.J. Stephen, Phys.\ Rev.\ {\bf 140},
A1197 (1965).
\bibitem{Blatt}G. Blatter, M.V. Feigel'man, V.B. Geshkenbein, A.I.
Larkin, and V.M. Vinokur, Rev.\ Mod.\ Phys.\ {\bf 66}, 1125 (1994);
C. de Morais Smith, B. Ivlev, and G. Blatter, Phys.\ Rev.\ B
{\bf 52}, 10~581 (1995).
\bibitem{FOG}A.T. Fiory, A.F. Hebard, P.M. Mankiewich, and R.E.
Howard, Phys.\ Rev.\ Lett.\ {\bf 61}, 1419 (1988);
S.B. Ota, R.A. Rose, B. Jayaram, P.A.J. de Groot, and
P.C. Lanchester, Physica C {\bf 157}, 520 (1989);
A. Gupta, P. Esquinazi, H.F. Braun, W. Gerh\"auser,
H.-W. Neu\-m\"ul\-ler, K. Heine, and J. Tenbrink, Europhys.\ Lett.\
{\bf 10}, 663 (1989).
\bibitem{Fish80}D.S. Fisher, Phys.\ Rev. B {\bf 22}, 1190 (1980).
\bibitem{Grish}A.M. Grishin, Y.M. Nikolaenko, A.V. Zinovuk,
B.Y. Vengalis, and A. Flodstr\"om, Fiz.\ Nizk.\ Temp.\ {\bf 19}, 42
(1993) [Low Temp.\ Phys.\ {\bf 19}, 30 (1993)].
\bibitem{Mooij}J.E. Mooij, in {\it Percolation, Localization, and
Superconductivity}, edited by A.M. Goldman and S.A. Wolf (Plenum,
New York, 1984), p.~325.
\bibitem{Mart}S. Martin, A.T. Fiory, R.M. Fleming, G.P. Espinosa,
and A.S. Cooper, Phys.\ Rev.\ Lett.\ {\bf 62}, 677 (1989).
\bibitem{VC}R.F. Voss and J. Clarke, Phys.\ Rev.\ B {\bf 13}, 556
(1976).
\end{references}
\end{document}